\documentclass[twocolumn]{aastex62}

\usepackage{newunicodechar}
\DeclareRobustCommand{\okina}{%
  \raisebox{\dimexpr\fontcharht\font`A-\height}{%
    \scalebox{0.8}{`}%
  }%
}
\newunicodechar{ʻ}{\okina}

\graphicspath{{./}{figures/}}

\shorttitle{Blastberries}
\shortauthors{Kong et al.}

\begin{document}

\title{
BlastBerries: How Supernovae Affect Lyman Continuum Escape Fractions and Ionizing Photon Production in Local Analogs of High-Redshift Galaxies}

\correspondingauthor{Miranda Y.\ Kong}
\email{ykong2@hawaii.edu}

\author[0000-0002-0786-7307]{Miranda Y. Kong}
\affiliation{Institute for Astronomy, University of Hawai\okina i, 2680 Woodlawn Drive, Honolulu, HI 96822, USA}

\author[0000-0002-6230-0151]{David O. Jones}
\affiliation{Institute for Astronomy, University of Hawai'i, 640 N Aʻohoku Pl, Hilo, HI 96720, USA}

\author[0000-0003-4761-2197]{Nicole E.\ Drakos}
\affiliation{University of Hawaiʻi at Hilo, 200 W Kawili St, Hilo, HI 96720, USA}

\author[0000-0002-9226-5350]{Sangeeta Malhotra}
\affiliation{Astrophysics Division, NASA Goddard Space Flight Center, Greenbelt, MD 20771, USA}

\author[0000-0001-9298-3523]{Kartheik Iyer}
\affiliation{Flatiron Institute, 162 5th Avenue. New York, NY 10010, USA}

\author[0000-0002-1428-7036]{Brian C.\ Lemaux}
\affiliation{Gemini Observatory, NSF NOIRLab, 670 N. Aʻohoku Place, Hilo, Hawaiʻi, 96720, USA}
\affiliation{Department of Physics and Astronomy, University of California, Davis, One Shields Ave., Davis, CA 95616, USA}

\author[0000-0003-3997-5705]{Rohan P.\ Naidu}
\affiliation{MIT Kavli Institute for Astrophysics and Space Research, 77 Massachusetts Avenue, Cambridge, MA 02139, USA}

\author[0000-0001-5486-2747]{Thomas de Boer}
\affiliation{Institute for Astronomy, University of Hawai\okina i, 2680 Woodlawn Drive, Honolulu, HI 96822, USA}

\author[0000-0001-6965-7789]{Ken C.\ Chambers}
\affiliation{Institute for Astronomy, University of Hawai\okina i, 2680 Woodlawn Drive, Honolulu, HI 96822, USA}

\author{John Fairlamb}
\affiliation{Institute for Astronomy, University of Hawai\okina i, 2680 Woodlawn Drive, Honolulu, HI 96822, USA}

\author[0000-0003-3953-9532]{Willem~B.~Hoogendam}
\altaffiliation{NSF Graduate Research Fellow}
\affiliation{Institute for Astronomy, University of Hawai\okina i, 2680 Woodlawn Drive, Honolulu, HI 96822, USA}

\author[0000-0003-1059-9603]{Mark E.\ Huber}
\affiliation{Institute for Astronomy, University of Hawai\okina i, 2680 Woodlawn Drive, Honolulu, HI 96822, USA}

\author[0000-0002-7272-5129]{Chien-Cheng Lin}
\affiliation{Institute for Astronomy, University of Hawai\okina i, 2680 Woodlawn Drive, Honolulu, HI 96822, USA}

\author[0000-0002-9438-3617]{Thomas Bernard Lowe}
\affiliation{Institute for Astronomy, University of Hawai\okina i, 2680 Woodlawn Drive, Honolulu, HI 96822, USA}

\author[0000-0002-7965-2815]{Eugene A.\ Magnier}
\affiliation{Institute for Astronomy, University of Hawai\okina i, 2680 Woodlawn Drive, Honolulu, HI 96822, USA}

\author{Paloma Mínguez}
\affiliation{Institute for Astronomy, University of Hawai\okina i, 2680 Woodlawn Drive, Honolulu, HI 96822, USA}

\author[0000-0002-6639-6533]{Gregory S.~H.~Paek}  
\affiliation{Institute for Astronomy, University of Hawai\okina i, 2680 Woodlawn Drive, Honolulu, HI 96822, USA}

\author{Angie Schultz}
\affiliation{Institute for Astronomy, University of Hawai\okina i, 2680 Woodlawn Drive, Honolulu, HI 96822, USA}

\author[0000-0002-1341-0952]{Richard J.\ Wainscoat}
\affiliation{Institute for Astronomy, University of Hawai\okina i, 2680 Woodlawn Drive, Honolulu, HI 96822, USA}



\begin{abstract}

While compact, star-forming galaxies are believed to play a key role in cosmic reionization, the physical mechanisms enabling the escape of ionizing photons through the galactic interstellar medium remain unclear. Supernova (SN) feedback is one possible mechanism for clearing neutral gas channels to allow the escape of Lyman continuum photons. Here, we use SN discoveries in low-redshift analogs of high-redshift star-forming galaxies --- Green Pea galaxies and their even lower-redshift counterparts, Blueberry (BB) galaxies --- to understand how SNe shape the properties of their host galaxies at high redshifts. We cross-match $1242$ BB galaxies with transient discovery reports and identify 11 SNe, ten of which are likely core-collapse SNe, and compare their hosts to the larger BB population. We find that SN-hosting BBs exhibit elevated star formation rates, burstier star formation histories within the last $\sim50$ Myr, and higher stellar masses. We estimate the occurrence rates of SNe in BB galaxies, finding that the SN rate may be slightly suppressed in BBs compared to field galaxies of similar mass, but we are unable to fully control for observational uncertainties.
Finally, SN hosts show bluer UV slopes than non-host BB galaxies at 2.1$\sigma$ significance and lower ionizing photon production efficiency at 7.9$\sigma$ significance; the former result offers modest support for the hypothesis that SN-driven feedback plays a role in facilitating the
escape of ionizing photons, while the latter may imply that SN-driven quenching decreases the rate of ionizing photon production in compact star-forming galaxies during the epoch of reionization.


\end{abstract}




\section{Introduction} \label{sec:intro}

Between redshifts $6 \lesssim z \lesssim 12$, the Universe underwent a major phase transition from predominantly neutral to fully ionized gas, marking the end of the cosmic dark ages \citep{fan_observational_2006,schroeder_evidence_2013,hinshaw_nine-year_2013,mcgreer_model-independent_2015,robertson_cosmic_2015,planck_collaboration_planck_2020}. Called the Epoch of Reionization (EoR), this was a critical period in cosmic evolution during which the first galaxies are believed to have been responsible for emitting ionizing radiation and ionizing the neutral intergalactic medium (IGM). However, the mechanisms and topology of reionization remain uncertain, with ongoing debate over the dominant ionizing sources \citep{madau_radiative_1999,barkana_beginning_2001,furlanetto_growth_2004,dayal_early_2018}, the timing and duration of reionization \citep{fan_constraining_2006,becker_evidence_2015}, and whether the process proceeded in an ``inside-out" or ``outside-in" manner: two contrasting topologies of ionization that differ in whether reionization starts in the densest regions and then expands outward into the low density IGM, or whether it starts in low density voids where recombination rates are low \citep{furlanetto_cosmology_2006,zaroubi_epoch_2013,kimm_escape_2014,madau_cosmic_2015,dayal_early_2018,robertson_galaxy_2022}.

Although the exact sources of ionizing photons are still an open question, radiation from low-mass, compact, star-forming galaxies (SFGs) is a popular candidate for satisfying the reionization photon budget and completing reionization by $z\sim6$ \citep{richards_sloan_2006,madau_cosmic_2014,robertson_cosmic_2015,robertson_galaxy_2022}. Young, hot stars emit Lyman-continuum (LyC) photons with $\lambda<912$\AA, and therefore enough energy to ionize hydrogen; these photons can escape into the IGM and ionize the surrounding neutral medium. Galaxies that leak a significant fraction of their ionizing photons are, therefore, strong contenders for dominating the ionizing photon budget, as their LyC emission carves out bubbles of ionized hydrogen in the IGM \citep{finkelstein_evolution_2015,sharma_brighter_2016}.

Recent theoretical frameworks have proposed various scenarios by which these high-$z$ galaxies might have contributed to cosmic reionization. These scenarios differ in their predictions of which galaxies are most responsible for reionization, and have consequences for the resulting topology of the IGM at high redshift.
Two contrasting paradigms include the so-called ``democratic" and ``oligarchic" models. In the democratic model, either numerous low-mass faint galaxies with modest escape fractions ($f_{\rm esc}$ $\sim10-20\%$) dominate the ionizing photon budget \citep{razoumov_ionizing_2010,finkelstein_evolution_2015,finkelstein_conditions_2019,mascia_new_2024,rinaldi_midis_2024}, or ultra-faint proto-galaxies leak an even larger fraction ($f_{\rm esc}$ $\sim30-40\%$) of their ionizing photons \citep{paardekooper_first_2013,paardekooper_first_2015}. Alternatively, in the ``oligarchic" model, a smaller population of relatively massive early-Universe galaxies --- $M_{\ast} \gtrsim 10^{8.5}~M_{\rm \odot}$ --- with high $f_{\rm esc}$ drive the majority of reionization by carving out large ionizing bubbles \citep{sharma_brighter_2016,naidu_rapid_2020}. 

Additional scenarios have also been proposed, including (1) AGN feedback models, where quasars and early AGNs are considered a non-negligible source of ionizing photons \citep{madau_cosmic_2024}, and can constitute $\sim10\%$ of the ionizing photon budget \citep[e.g.,][]{dayal_reionization_2020}; (2) environment-dependent escape fractions, where tidal forces and mergers can significantly change the structure and column density of the interstellar medium \citep[ISM;][]{madau_cosmic_2015,dayal_reionization_2020,lewis_dustier_2023,jiang_agns_2025}; and (3) intermediate-mass galaxies that sit between the extreme scenarios, corresponding to a peak far-ultraviolet absolute magnitude ($M_{FUV}$) at which galaxies contribute the most, with contributions decreasing toward both fainter and brighter $M_{FUV}$ \citep{ma_no_2020,matthee_resolving_2022,lin_empirical_2024}. It is worth noting that although these models primarily focus on the properties of individual galaxies, protoclusters and groups can also act in tandem to carve out bigger ionized bubbles or induce ionizing AGN activity \citep{tilvi_onset_2020,shah_enhanced_2024}.

The LyC escape fractions in high-$z$ SFGs also depend on halo mass, column density, dust production, and ISM geometry \citep{shapley_direct_2006,wise_ionizing_2009,wise_birth_2014}. The ``democratic" scenarios posit that lower-mass galaxies (and thus lower-mass dark matter halos) should have higher Lyman escape fractions due to low gas covering fractions and shallower potential wells, making it easier for ionizing photons to penetrate the ISM. In comparison, massive galaxies and their deeper potential wells form high-density gas reservoirs with column densities that render them effectively opaque to LyC photons \citep[e.g. $M_{\rm halo}>10^8-10^9M_{\rm \odot}$,][]{gnedin_escape_2008,paardekooper_first_2013,kimm_escape_2014,finkelstein_conditions_2019}.

However, mechanical energy injections from feedback can change the galaxy topology. Feedback processes can clear or significantly lower the ISM density, increasing $f_{\rm esc}$ \citep{heckman_escape_2001,clarke_galactic_2002,keller_uncertainties_2022,flury_low-redshift_2025}. Effects from stellar feedback, AGN feedback, SN feedback, and resulting interstellar winds are all proposed as critical sources of mechanical energy that can clear out gas and dust channels through the ISM \citep{heckman_extreme_2011,amorin_complex_2012,trebitsch_fluctuating_2017,nelson_first_2019,flury_low-redshift_I_2022,llerena_ionized_2023,jiang_agns_2025}. The timeline of these feedback effects is considered to be around $2-3$~Myr, though the process of clearing out channels can extend up to $10$~Myr \citep{naidu_synchrony_2022}. These time scales, however, remain mainly theoretical and are yet to be directly tested.

Currently, the level at which such feedback processes contribute to reionization is poorly constrained. Studying galaxies during the EoR is difficult because they are so distant. Studying SNe at these redshifts is even more difficult; even with the {\it James Webb Space Telescope} ({\it JWST}), the highest redshift SN candidate to date is at $z = 5.274$ --- several hundred Myrs after the canonical end of reionization --- and has only a single epoch of observation \citep{decoursey_first_2025}. The highest redshift SN confirmed with spectroscopy is at $z<4$ \citep{decoursey_jades_2024,coulter_discovery_2025}. There have also been observations of gamma-ray bursts following a SN candidate at $z\simeq7.3$, deep in the era of reionization, though without direct SN observations \citep{cordier_svom_2025,levan_jwst_2025}.

Fortunately, a population of galaxies at low redshifts has attributes similar to SFGs during the EoR. High-redshift SFGs are usually compact in size, with low stellar masses, dust extinctions, metallicities, oxygen abundances, and blue UV continuum slopes, but with high emission-line equivalent widths of high-ionization tracers like [O~III], and high specific star-formation rates \citep[SFRs;][]{yang_ly_2017,izotov_universality_2015,izotov_low-redshift_2021}. A population of galaxies in the low-redshift regime, named Green Pea Galaxies (GP) after their optical color in SDSS filters \citep{cardamone_galaxy_2009}, match these attributes and are commonly considered to be local analogs of high-redshift SFGs \citep{yang_green_2016,kim_compact_2021,schaerer_first_2022}. 

GPs owe their optical colors to extremely high emission line equivalent widths (EWs), especially [O~III] $\lambda$5007, and high contrast between their lines and continua. Their high line equivalent widths result from high star formation and the lack of a prominent older stellar population, which together sustain a hard ionizing spectrum and elevated electron temperatures; this leads to particularly strong [O~III] $\lambda$5007 emission and other high-ionization lines alike \citep[e.g., Ne~III;][]{berg_direct_2012,jaskot_origin_2014,senchyna_ultraviolet_2017}. At the lowest redshifts --- here, we choose $z<0.12$ following color selection criteria from \citealp{liu_strong_2022} --- we refer to these galaxies as ``Blueberries" (BBs)\footnote{We note that the exact redshift range used to define ``Blueberry" galaxies varies in the literature.}, as their [O~III] line emission dominates the $g$-band \citep{yang_green_2016,yang_ly_2017}. 

Although observations are not yet powerful enough to detect SNe in EoR galaxies, discovering them in GP and BB galaxies is much more tractable.  In recent years, high-cadence all-sky surveys such as the Zwicky Transient Facility \citep[ZTF,][]{bellm_zwicky_2019}, the Asteroid Terrestrial-impact Last Alert System \citep[ATLAS,][]{tonry_atlas_2018}, the ALL-Sky Automated Survey for Supernovae \citep[ASAS-SN,][]{shappee_all_2014} and the Panoramic Survey Telescope and Rapid Response System \citep[Pan-STARRS,][]{chambers_pan-starrs1_2016}, have begun discovering tens of thousands of SNe, particularly at low redshifts ($z < 0.1$) where BB galaxies reside.
With future surveys such as the Vera C.\ Rubin Observatory's Legacy Survey of Space and Time \citep[LSST;][]{brough_vera_2020} and the {\it Roman Space Telescope} high-latitude time-domain survey \citep{rose_reference_2021,rose_hourglass_2025,observations_time_allocation_committee_roman_2025}, the size of the SN sample will continue to increase through ongoing discoveries, follow-up observations, and classifications.

In this paper, we search for recent SN explosions occurring in two samples of BB galaxies \citep{jiang_direct_2019,liu_strong_2022}.
Our analysis is predicated on the fact that galaxies in which a SN has been directly observed are likely to have higher SN explosion frequencies overall compared to non-host galaxies, while the same types of SNe are found more often in multi-SN hosts \citep{anderson_multiplicity_2013}. This implies that past SN explosions will have shaped the ISM in these galaxies to a greater degree than non-SN hosts; this is particularly true for CC SNe, whose rates follow the bursty SFHs of these BB galaxies.  SN\,II progenitors, for example, are preceded by the explosions of the more massive SNe\,Ib and Ic progenitors in the Myr prior to their own terminal explosions. 

After selection cuts, we identify 11 SNe in 1242 $z < 0.12$ BB galaxies. Using this sample, we compare SN-host galaxy masses, SFRs, ultraviolet (UV) slopes, and other derived quantities to the larger BB sample and field galaxies as a whole.  We infer the relative SN rate for these BB host galaxies, which has implications for how much mechanical energy SNe are injecting into the neutral ISM at high redshifts, and use these derived quantities to better understand the role of SN explosions in allowing LyC escape during the EoR.

In Section~\ref{sec:methods}, we present the BB galaxy samples and SN light curves. We measure relative SN rates and galaxy properties in Sections~\ref{sec: rates} and \ref{sec: host properties}, respectively. Finally, we discuss implications for cosmic reionization in Section~\ref{sec:discussion}. Throughout this study, we use a flat $\Lambda$CDM cosmology with H$_0 = {\rm 70~km~s^{-1}~Mpc^{-1}}$ and $\Omega_m=0.3$ for all luminosities and absolute magnitudes, with magnitudes given in the AB system. All SED fitting and scaling relations are based on the Chabrier initial mass function \citep{chabrier_galactic_2003}.

\section{Data and Methods} \label{sec:methods}

\begin{figure*}
    \centering
    \includegraphics[width=0.75\linewidth]{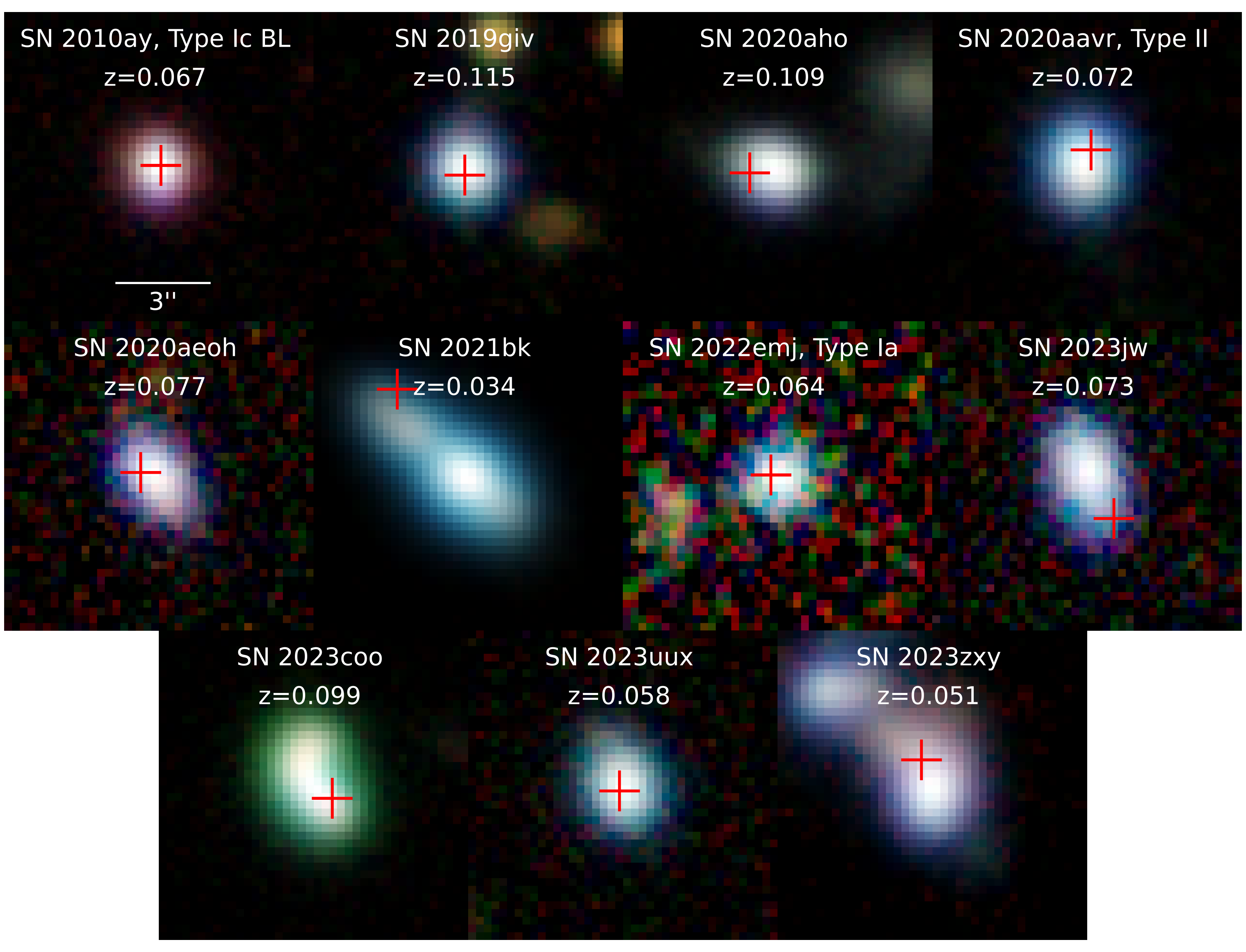}
    \caption{Three-color stamp images of the BB SN hosts in our sample, using \textit{grz} filters from the DESI Legacy Imaging Survey \citep{dey_overview_2019,levi_dark_2019}. The images are centered on the host galaxies, while the red crosses mark the coordinates of each SN. Redshifts are derived from SDSS catalogs, which include spectroscopically derived parameters for each host galaxy \citep{york_sloan_2000}. The scale bar is given in angular size in the top left panel, where 3" corresponds to $\sim3$ kpc at $z=0.05$ and $\sim5.5$ kpc at $z=0.1$. Classification information is from the Transient Name Server (TNS) when available; some SNe in our sample were never spectroscopically classified.}
    \label{fig:stamps}
\end{figure*}

\subsection{Green Pea Galaxy Selection}
\label{subsec:data}
We use two publicly available GP catalogs for a total sample of 2382 galaxies. The first catalog is selected from the Large Sky Area Multi-Object Fiber Spectroscopic Telescope \citep[LAMOST,][]{luo_design_2004}, and the second is from the Sloan Digital Sky Survey \citep[SDSS,][]{albareti_13th_2017}. We describe the catalogs below, but for more data details and selection methods, see the catalog papers, \citet{jiang_direct_2019} and \citet{liu_strong_2022}, respectively.  We note that we adopt the \citet{jiang_direct_2019} and \citet{liu_strong_2022} GP identification criteria for each respective sample in this study, and do not attempt to re-evaluate compactness or color selections to better match EoR galaxies.

LAMOST has carried out spectral observations for millions of stars and galaxies since its commissioning.  Its data release (DR) 9 includes 2309 spectra of candidate GP galaxies in the northern sky. From this initial sample, \citet{liu_strong_2022} selected 1547 GP galaxies with strong [OIII] $\lambda$5007 emission lines and high equivalent widths, compact physical sizes, and in the redshift range $0<z<0.722$.

\begin{table*}
\centering
\begin{tabular}{l  r  r  r r }
 \hline
 Survey  & Total Sample & $z<0.12$ Sample & SN Matches & Reference\\
 \hline
 LAMOST GP& $1694$ & $791$ & $4$ &\citet{liu_strong_2022}\\
 SDSS GP & $1004$ & $482$ & $7$ & \citet{jiang_direct_2019}\\
 SDSS Field & $778\,914$ & $311\,769$ & $7691$ &\citet{albareti_13th_2017}\\
 \hline
\end{tabular}
\caption{Sizes of the LAMOST and SDSS GP samples and the SDSS field-galaxy sample. We include the size of the full sample of galaxies and the filtered $z<0.12$ subsets, which we address as BB galaxies and their field counterparts in this study.}
\label{table: Sample}
\end{table*}

To select GPs, \citet{liu_strong_2022} used a two-fold strategy for the color selection depending on morphology and redshift constraints. For sources in the $z<0.12$ range, which is the focus of this work (see Section~\ref{sec:snsample}), they applied loose color criteria with $u-g\leqslant0.3$, $r-g\leqslant0.1$, and $g-i \leqslant 0.7$ for extended sources, or strict color criteria with $u-g\leqslant0.5$ and $r-g\leqslant0.5$ when the source had no morphological constraints.

\citet{liu_strong_2022} placed additional constraints on the compactness and line strengths of the color-selected sample, choosing only galaxies with $r$-band radii of $r<5$\arcsec \citep[at the typical redshift of peas, this corresponds to an upper limit of $\sim5$ kpc in physical half-light radius][]{cardamone_galaxy_2009} and an [OIII] $\lambda$5007 line flux above $3\times10^{17}\, {\rm erg \,s^{-1}\,cm^{-2}}$.

The second sample comes from a collection of optical spectra of SDSS galaxies from the Baryon Oscillation Spectroscopic Survey \citep[BOSS;][]{ahn_tenth_2014} with corresponding reprocessed Sloan Digital Sky Survey (SDSS; \citealp{york_sloan_2000}) legacy survey images. \citet{jiang_direct_2019} selected 835 GP galaxies in the redshift range $0.011<z<0.411$. The selection criteria requires the galaxies to be spectroscopically classified as a ``star-forming", ``starburst", or ``NULL" galaxy subclass (the latter classification denoting that it has no distinct features of any subclass), but not ``AGN". The selection also requires that the [O~III] $\lambda$5007 and $H\beta$ lines are detected with a signal-to-noise ratio (S/N) $>5$ and have high rest-frame equivalent widths of ${\rm EW([O~III]}\lambda5007)>300$\AA\ or EW(H$\beta)>100$\AA, and must be spatially compact with an $r$-band radius smaller than 3\arcsec. They include only those sources that are classified as star-forming galaxies by BPT diagrams, a common tool for classifying galaxies by distinguishing their gas ionization mechanisms \citep{baldwin_classification_1981}. Additionally, the sample chose galaxies with a ${\rm S/N}>3$ for [O~III] $\lambda$4363, though due to its close correlation with [O~III] $\lambda$5007, the effect of this filter is minimal.
 
We note there are two main differences between the \citet{liu_strong_2022} and \citet{jiang_direct_2019} selection criteria.  While \citet{liu_strong_2022} apply strict color criteria based on imaging, the \citet{jiang_direct_2019} sample is primarily determined by line strengths and EWs. \citet{jiang_direct_2019} sets thresholds on both [OIII] and $H\beta$, while \citet{liu_strong_2022} uses only [OIII] $\lambda$5007. \citet{jiang_direct_2019} also applies a tighter cut on spatial compactness of the $r$-band radius $<3$\arcsec.  Additionally, upon visual inspection of a subsample in the optical band, we find that some \citet{liu_strong_2022} GP candidates are bright star-forming knots embedded in larger extended galaxies; however, we found that $>$99\% are bona fide GPs, and that this contamination will have a statistically insignificant effect on our SN occurrence rate measurements (Section \ref{sec: rates}).

Throughout, we use the full SDSS spectroscopic catalog from DR13 as a comparison sample. We filter the catalog to include only ``galaxy" type objects, and limit to $z<0.12$ to align with the BB sample.

\subsubsection{Supernova Search}
\label{sec:snsample}
For every galaxy in the BB sample, we perform an initial search for SNe discovered within 30\arcsec\ of their host centers, a radius $\gtrsim6-10$ times the size of a typical BB galaxy that was chosen to ensure that all SNe associated with a given host can be found. We search through all SNe that have been publicly reported to the International Astronomical Union (via the Transient Name Server\footnote{\url{https://www.wis-tns.org/}.}). Once we find a possible match, we visually inspect the results to see if the SN is coincident with the galaxy. All of our bona fide SNe were found within $\sim$2\arcsec\ of the host-galaxy center, demonstrating that our conservative initial 30\arcsec\ matching threshold is sufficient for compact galaxies like BBs.

We find 11 SN candidates within close proximity to, and likely hosted by, one of the BB galaxies; an additional 3 are found in the higher-redshift GP sample and are later rejected from our sample due to quality cuts (see below).   The sizes of our samples and the numbers of identified SN hosts are listed in Table \ref{table: Sample}. All but one of these SNe occurred during the last seven years, due to the recent increase in SN discoveries. The exception is SN~2010ay, a Type Ic-BL SN discovered by Pan-STARRS and published in \citet{sanders_sn_2012}.  We obtained light curve data from the ZTF, ATLAS, and Pan-STARRS surveys to confirm that the candidates were real SNe and describe these data below.

First, ZTF is a wide-field optical time domain survey that scans the entire Northern sky in the $g$, $r$, and $i$ filters. We obtained ZTF data using the publicly available ZTF forced photometry service \citep{masci_zwicky_2019}, which allows users to request forced-photometry lightcurves at any coordinate on the sky.

ATLAS consists of four telescopes in Hawai'i (Maunaloa and Haleakal\=a), Chile, and South Africa, which scan the whole sky several times each night in the cyan and orange broadband filters (equivalent to $g+r$ and $r+i$, approximately). We obtained these data using their forced-photometry service \citep{shingles_release_2021}.\footnote{\url{https://fallingstar-data.com/forcedphot/}.}

The Pan-STARRS telescopes also provide wide-field imaging of the night sky in $griz$ filters.  Photometry is provided by the Pan-STARRS image processing pipeline (IPP) at the University of Hawaiʻi; the IPP performs data processing, calibration, difference imaging, and photometry for all Pan-STARRS data \citep{flewelling_pan-starrs1_2020,magnier_pan-starrs_2020}.  Data are passed to a version of the Transient Science Server \citep{smith_design_2020}, which carries out searches for new SNe as part of the Pan-STARRS Survey for Transients \citep[PSST,][]{huber_pan-starrs_2015}.  Once a candidate SN is discovered, forced photometry is automatically carried out at the location of the SN.

\begin{figure*}
    \centering
    \includegraphics[width=0.8\linewidth]{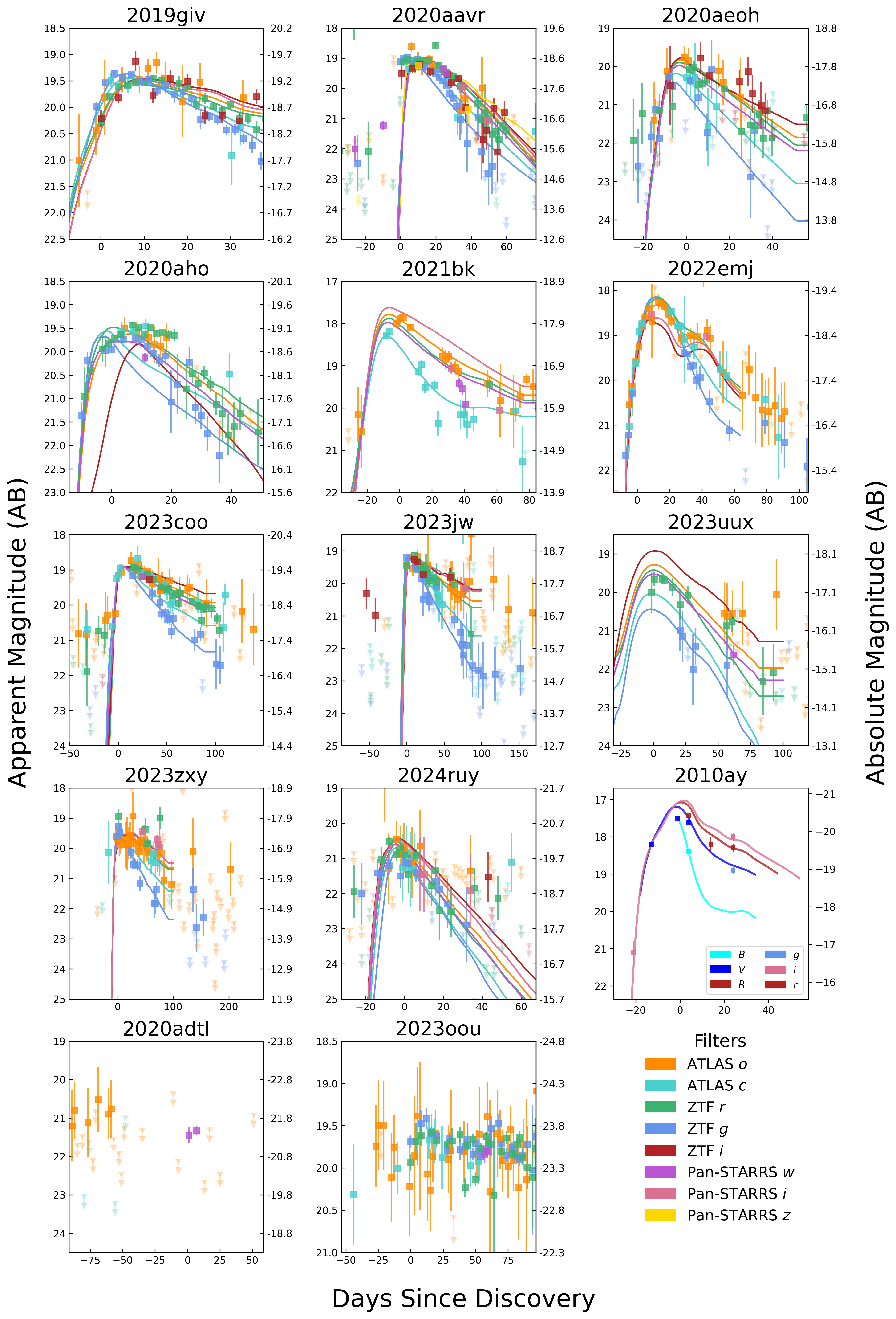}
    \caption{ZTF, ATLAS, and Pan-STARRS light curves for all SNe in our matched sample, with SN~2010ay from \citet{sanders_sn_2012} including synthetic photometry from Gemini observations. SN candidates 2024ruy, 2020adtl, and 2023oou are excluded from our analysis. We show the best-fit {\tt STARDUST2} model for each bona fide SN \citep{rodney_type_2014}. }
    \label{fig:LC}
\end{figure*}

We constructed light curves using the available photometry with all 14 SNe, as shown in Figure~\ref{fig:LC}. We note that SN 2010ay predates the start of the ZTF and ATLAS surveys, hence we were only able to retrieve Pan-STARRS photometry; we additionally compiled synthetic photometry and Gemini/GMOS data from \citet{sanders_sn_2012} as part of Figure~\ref{fig:LC}. 

Upon visual inspection and redshift filtering, we remove three of the 14 candidates from the sample.
Two (2020adtl and 2023oou) are identified as QSO objects by SDSS; they also do not appear to have a clear monotonic rise or fall in their light curves, and therefore are likely not real SNe. Furthermore, they both have redshifts of $z>0.6$; typical SNe at these redshifts would be well below the detection threshold of ATLAS, ZTF, or Pan-STARRS. One additional SN candidate, SN~2024ruy, has a redshift of $0.26$, higher than the rest of the SNe hosts and the local BB population; it is removed to limit our BB sample to $z<0.12$, which also preserves consistency in color selection with the LAMOST sample.

After selection cuts, our final sample includes four SNe in the LAMOST galaxies, and seven in the SDSS galaxies, for a total of 11 SNe. Detailed information for these SNe and their hosts is listed in Table~\ref{table: SN Basics}. Three SNe were classified spectroscopically: 2010ay as Type Ic-BL, 2020aavr as Type II, and 2022emj as Type Ia. These 11 SNe are considered to be our final sample of BB-hosted SNe and the main subject of study for this paper.

\begin{table*}
\centering
\begin{tabular}{l r r r r r }
 \hline
 SN ID & Host Survey & Host ID & Host z & Spec.\ Class.\ & Abs.\ Mag.\\
 \hline
 2010ay  & LAMOST  & SDSS J123527.19+270402.7 & $0.0671$ & Ic-BL & $-20.2$\\
 2023uux &  LAMOST  & SDSS J104040.26+055405.9 & $0.0577$ & \nodata & $-17.4$\\
 2020aeoh & LAMOST & SDSS J100741.62+082536.4 & $0.0767$ & \nodata & $-18.2$\\
 2023jw & LAMOST & SDSS J123154.45+313707.6 & $0.0730$ & \nodata & $-18.6$\\
 2020aavr & SDSS & SDSS J085949.12+380632.1 & $0.0725$ & II & $-18.6$\\
 2020aho & SDSS & SDSS J082722.57+202612.8 & $0.1086$ & \nodata & $-19.2$\\
 2023zxy & SDSS & SDSS J122611.15+532602.0 & $0.0514$ & \nodata & $-17.9$\\
 2022emj & SDSS & SDSS J121234.81+463541.2 & $0.0643$ & Ia & $-19.3$\\
 2019giv & SDSS & SDSS J154659.01+325632.2 & $0.1148$ & \nodata & $-19.5$ \\
 2021bk & SDSS & SDSS J135950.91+572622.9 & $0.0338$ & \nodata & $-18.1$\\
 2023coo & SDSS & SDSS J112145.16$-$000121.2 & $0.0990$ & \nodata & $-19.7$\\
 \hline
\end{tabular}
\caption{BB-hosted SNe, including their respective hosts, redshifts, spectroscopic classifications, and approximate absolute magnitudes at peak. The absolute magnitude for 2010ay is taken from \citet{sanders_sn_2012}, while all others are measured from our lightcurve data. SN~2010ay, 2020aavr, and 2022emj are spectroscopically classified. The unclassified SNe are likely core-collapse SNe, which we determine using photometric classification in Section~\ref{subsec:SN class}.}
\label{table: SN Basics}
\end{table*}

We also cross-match TNS objects to field galaxies to use as a comparison sample, first filtering the field galaxies to match the redshift range of the BB SN hosts ($z < 0.12$).  This gives the field-galaxy sample comparable selection effects (i.e., Malmquist bias) to the BB sample.  For these galaxies, we conservatively perform a cone search for SNe within a 75\arcsec\ radius of each host galaxy center.  We then keep only those SNe for which the galaxy is the most probable host, as determined by the GHOST algorithm \citep{gagliano_ghost_2021}; GHOST finds the most likely host galaxy for each SN by using a gradient ascent algorithm trained on the Pan-STARRS photometric catalog.  This cross match yields a total of $7691$ SNe from our sample of $311\,769$ SDSS field galaxies.

\subsection{Supernova Classifications}
\label{subsec:SN class}

For those SNe lacking spectroscopic classifications, we use their peak luminosities, model fits with the SALT3 SN\,Ia standardization model \citep{kenworthy_salt3_2021}, and the {\tt STARDUST2} classifier \citep{rodney_type_2014} to derive approximate SN types.
Our primary classifier is {\tt STARDUST2}, a Bayesian classifier originally built for high-redshift SNe discovered by the {\it Hubble Space Telescope} ({\it HST}), which compares the data to simulated SN light curves from 27 Type II and 15 Type Ib/c templates, as well as the SALT3 SN\,Ia model \citep{kenworthy_salt3_2021,pierel_salt3-nir_2022}.  Realistic luminosity and dust distributions are applied to the templates, with additional details provided in \citet{rodney_type_2014,decoursey_jades_2024}.  The number of simulated samples for each SN type is scaled by the volumetric rates from \citet{dilday_measurement_2008} and \citet{li_nearby_2011}.  From these simulations, {\tt STARDUST2} then marginalizes over the probability that a SN is Type Ia, Ib/c, or II.  For samples with limited data, such as ours, the strength of {\tt STARDUST} is that it does not require a large sample of simulated or real light curves to train a machine-learning algorithm.

From just peak magnitudes alone, we find that four of eight unclassified SNe are likely CC\,SNe; from \citet[their Table 1 at $z > 0.02$]{desai_supernova_2024}, the mean absolute magnitude of SNe\,Ia in a magnitude-limited sample is approximately $V \simeq -19.05\pm0.44$~mag.  This is strongly inconsistent with the observed luminosities of SNe\,2023uux and 2023zxy, which {\tt STARDUST2} classifies as SNe\,Ib/c and II, respectively. SNe\,2021bk and 2020aeoh have a $<$3\% chance of being SNe\,Ia; {\tt STARDUST2} classifies both as SNe\,Ib/c. 

For the remaining SNe, we use the SALT3 model \citep{kenworthy_salt3_2021} as implemented in {\tt sncosmo} \citep{barbary_sncosmo_2016} to determine whether the best-fit light curves are consistent with a SN\,Ia.  We find that SN~2019giv rises to peak approximately twice as fast ($\sim$8 days) as would be expected for its best-fit stretch; accordingly, {\tt STARDUST2} classifies it as a SN\,II.  SN\,2020aho's luminosity is consistent with a SN\,Ia, and {\tt STARDUST2} classifies it as a SN\,Ia, but the SALT3 fit shows inconsistent early versus late-time colors between data and model, as well as a prominent bump in the early-time $g$-band light curve that indicates shock breakout characteristic of a SN\,II (some SNe\,Ia have early time excesses, but typically they occur at several mag below peak; \citealp{hoogendam_out_2024}).  SN~2023jw is somewhat underluminous for a SN\,Ia, and its best-fit shape ($x_1$) and color ($c$) parameters are inconsistent with a SN\,Ia (too high-stretch and red, respectively); {\tt STARDUST2} classifies it as a SN\,II.  Lastly, SN\,2023coo declines too slowly to be a SN\,Ia, and {\tt STARDUST2} classifies it as another SN\,II.

Overall, we find that the SNe in our sample, with the exception of SN\,2022emj, are likely to be CC\,SNe.  This means that they should probe the star-forming environment of their host galaxies within the last $\sim$10-50~Myr \citep{zapartas_delay-time_2017}. We note that our sample has a high fraction of core-collapse SNe when compared to the field, though we do not further probe the characteristics of these SNe in this work.

\section{Relative Occurrence Rates of Supernovae in Blueberry Galaxies} \label{sec: rates}

Understanding how frequently SNe occur in different galaxy environments provides insight into the nature of their stellar populations and the feedback processes that regulate galaxy evolution. In particular, the rate of CC SNe is closely tied to recent star formation, reflecting the short lifetimes of massive progenitor stars \citep{mannucci_supernova_2005,kennicutt_star_2012,strolger_rate_2015}. To evaluate whether the compact star-forming environment of BB galaxies has a measurable effect on the SN rate, we estimate the relative frequency of CC\,SNe in 1242 BB galaxies (SDSS and LAMOST combined) and compare this sample to the redshift-matched field sample.

We first derive approximate host-galaxy masses and SFRs from scaling relations relying on optical and UV photometry.  These allow us to quickly measure SED properties across more than $10^6$ galaxies, providing a determination of SN rates from the field-galaxy sample as a function of those properties.  We obtain GALEX and SDSS photometry for the full BB and SDSS field-galaxy sample at $z < 0.12$ by querying their respective catalogs. All galaxies have spectroscopic redshifts for determining estimated distance moduli, used to measure absolute magnitudes.

We measure the mass using the scaling relation provided by \citet[their Equation 8]{taylor_galaxy_2011}, which is based on SDSS $gi$ magnitudes.
To validate these masses, we compare them to SDSS catalog mass measurements, which use the spectroscopically derived redshifts and $ugriz$ photometry \citep{maraston_stellar_2013}, for a subset of our sample.  This comparison yields a statistically significant correlation coefficient of $0.728$, a mean offset of $0.06$ dex, and a scatter of $0.63$ dex; scatter is dominated by the low-mass end, and is likely due to the breakdown of color--M/L calibrations in emission-line dominated SFGs \citep{roediger_uncertainties_2015,sorba_missing_2015}.
We also note that both mass measurements assume parametric star-formation histories, which can cause stellar masses to be underestimated by $\sim$0.1-0.2~dex \citep{leja_how_2019}.

We also specifically examine the low-mass range of this comparison, finding slightly higher scatter of $\sim$0.8~dex between $6 < \log(M_*/M_{\rm \odot}) < 7$.  Because BB~SNe have high [OIII] flux in the $g$ band, it is possible that this band is line-dominated; this effect would systematically reduce the \citet{taylor_galaxy_2011} mass estimates by changing the color correction term.  However, we see that the \citet{taylor_galaxy_2011} masses are generally higher (by $\sim$0.4~dex) than the SDSS spectroscopic measurements for these low-mass data (but with significant scatter).


Our SFR measurement uses the relationship from \citet[see also \citealp{rigault_confirmation_2015}]{salim_uv_2007} derived from GALEX {\it FUV} magnitudes \citep{martin_galaxy_2005}. The dust-corrected UV emission of galaxies redward of the LyC break directly traces the emission of relatively young stars, and is therefore a good tracer of SFR on $\sim$100~Myr timescales \citep{flores_velazquez_time-scales_2021}. We correct for Milky Way extinction using the \citet{schlegel_maps_1998} reddening maps, the \citet{schlafly_measuring_2011} dust temperature correction, and a reddening law of $R_V = 3.1$.  We approximately correct for the galaxy extinction, $A_{FUV}$, using {\it FUV}$-${\it NUV} colors following \citet{rigault_confirmation_2015}, and convert to {\it FUV} luminosity using the distance modulus. We then use the relationship from \citet{kennicutt_star_2012} to infer the time-delayed SFR from the dust-corrected {\it FUV} luminosities:
\begin{equation}
\label{eqn:SFR}
    \log \left(\frac{SFR} {M_{\rm \odot}\,{\rm yr}^{-1}}\right)  = 1.2\times10^{-28}L_{ \nu,FUV} \,{\rm (erg/s/Hz)}.
\end{equation}
We again compare these calculations to SDSS spectroscopically derived measurements, and find a statistically significant linear correlation with a coefficient of $0.76$ and a standard deviation of $0.44$~dex.  We note that the SDSS fiber diameter (2--3\arcsec) might not cover the full region of more extended field galaxies --- and preferentially include older stellar populations near the galaxy center --- and therefore underestimate their SFR, which is less of a concern for compact BB galaxies, resulting in subtle biases in this comparison.

\subsection{Results}

\begin{figure}
    \centering
    \includegraphics[width=\linewidth]{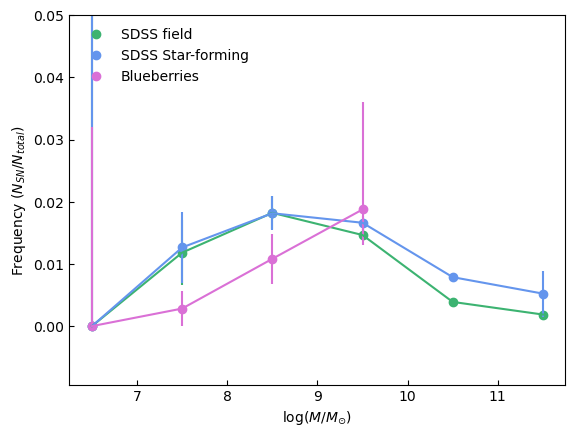}
    \caption{SN event frequency ($N_{\rm SN}/N_{\rm total}$) as a function of mass for the SDSS field galaxy sample (green), the color-filtered SFG sample (blue), and the BB sample (pink). Error bars are derived from Poisson statistics for galaxies in the given stellar mass bin. The BB sample shows a lower frequency of SN events.
    }
    \label{fig:rates}
\end{figure}

\begin{figure}
    \centering
    \includegraphics[width=\linewidth]{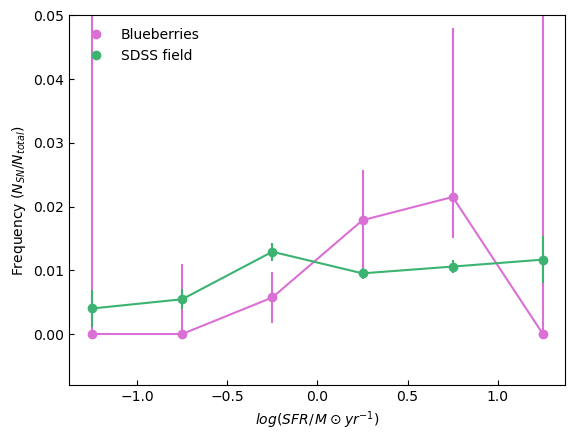}
    \caption{SN frequency for the SDSS field galaxy sample (green) and the BB sample (pink). Each sample is binned by FUV-calculated SFR, and frequency is calculated by $N_{{\rm SN}}/N_{{\rm total}}$ in each bin.  See the Figure \ref{fig:rates} caption for additional details.
    }
    \label{fig:rates_SFR}
\end{figure}

In Figures~\ref{fig:rates} and \ref{fig:rates_SFR}, we show the fraction of galaxies hosting SNe, for both BB and field galaxies, binned by mass and SFR respectively. 
Because we wish to specifically study the rate of CC\,SNe, we include only likely CC\,SNe in the BB sample (all SNe except for SN~2022emj). We also include only classified CC\,SNe in the field galaxy sample, but correct their measured rates by the overall classification fraction of SNe hosted by $z < 0.12$ SDSS galaxies ($\sim$25\%) so that we can fairly compare these relative rates to the photometrically classified BB SN sample. 

Compared to the full field galaxy sample at stellar masses $<$10~dex --- to limit to the mass range where BB galaxies are found --- BBs show lower SN frequency; approximately $0.9\pm0.3\%$ of BB galaxies in our sample were observed to host CC\,SNe, while $1.5\pm0.1\%$ of field galaxies hosted CC\,SNe after correcting for the classification incompleteness as described above.  The difference in frequency is $\Delta{f}=0.59\pm0.29\%$, significant at the $2.1\sigma$ level.  However, the field galaxies overall have higher masses than the BB sample; when we require a mass limit such that the median mass of the BB sample is the same as that of the field sample, we find a less confident $\Delta{f}=0.46\pm0.37\%$ (1.2$\sigma$).  There is also some potential for systematic bias in our mass calculations at the lowest masses, as discussed above, which could reduce the significance of this difference further.  

Additionally, the subset of galaxies for which GALEX {\it FUV} measurements are available show a slightly {\it higher} SN rate in the BB galaxies than the field sample ($1.0\pm0.05$\% for the field versus $1.1 \pm 0.3$\% for the BB sample).  However, because only 1/10 BB-hosted CC~SNe have unreliable GALEX {\it FUV} data, while 34\% of the field sample is missing {\it FUV} data, we consider this result to be dominated by statistical fluctuation.




In Figure \ref{fig:rates}, we show the relative frequency of SNe as a function of mass in BBs, all SDSS field galaxies, and field SFGs.  To isolate the SFG sample, we select galaxies with $u-r<2.3$~mag, which follows the bimodal distribution of the $u-r$ color in our sample. We note that this color cut excludes both quiescent galaxies and dusty star-forming galaxies (DSFGs), which are predominantly bright in the mid-to-far infrared.
\citep{blain_submillimeter_2002,chapman_redshift_2005,casey_dusty_2014,bussmann_hermes_2015}. This filter yields $139\,430$ SFGs.  We see largely consistent rates between all field galaxies and field SFGs at masses $<$10~dex, which is unsurprising given that most of these galaxies are star-forming; at higher masses, the difference between the overall field rates and the SFG rates reflect the higher fraction of passive galaxies at these masses, which do not host CC\,SNe.



 Both Figures \ref{fig:rates} and \ref{fig:rates_SFR} show that BB galaxies tend to have a lower rate of SN occurrence compared to the field galaxies in the low mass and low SFR bins (at $\sim$2$\sigma$ significance), while at high mass and SFR the rates are comparable.
In addition to the caveats above, however, we also note that SNe in BB galaxies may be systematically under-reported, which may be due to their host galaxies being mistaken for high-redshift AGN, and would cause their rate to be higher than calculated in this study. In future work, we plan to perform a more exhaustive forced-photometry search for previously undetected BB-hosted SNe, with surveys such as Pan-STARRS and ZTF.  We discuss the potential implications of low SN rates in BB galaxies in Section \ref{sec:discussion}.

\section{Host Properties} \label{sec: host properties}

The role of SNe in shaping galaxy-scale environments can be traced by the physical conditions of their host galaxies. In this section, we explore whether BB galaxies that host SNe differ systematically from the broader BB sample. In Section~\ref{subsec:properties}, we investigate the correlation between the presence of SNe and the galaxy stellar mass, SFR, and metallicity; in Section~\ref{subsec:UV slopes}, we dive deeper into the galaxies' UV slopes and the implied Lyman continuum escape fractions, while in Section~\ref{subsec:SFH} we measure the star formation histories of these galaxies.
\begin{figure*}[!]
    \centering
    \includegraphics[width=\textwidth]{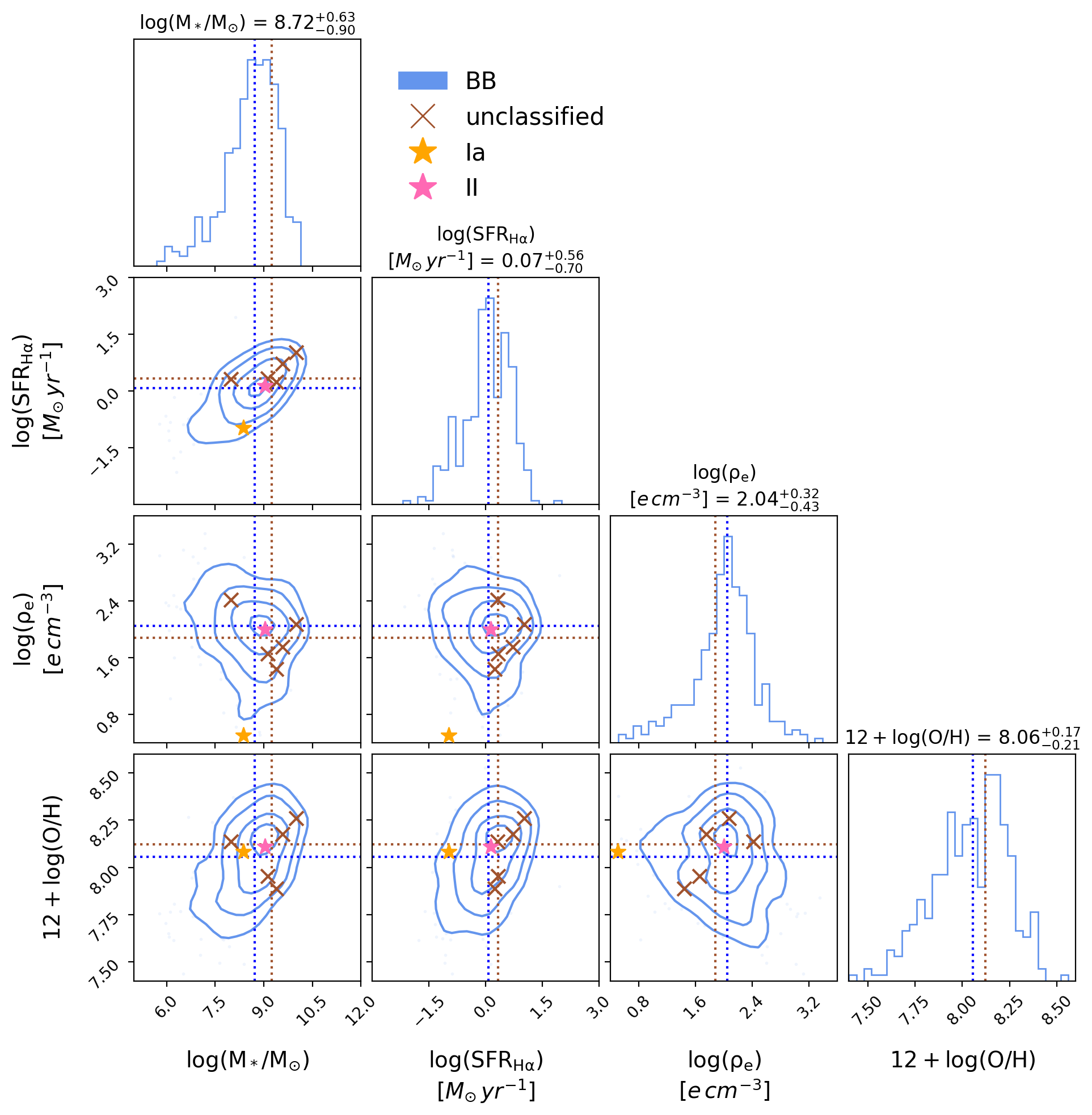}
    \caption{Corner plot showing galaxy properties of BB galaxies and BB SN hosts, measured from the SDSS spectroscopic pipeline. Properties include the H$\alpha$-inferred SFR, electron density, metallicity, and stellar mass. The contours displayed correspond to the $0.5, 1, 1.5, 2 \sigma$ levels. The unclassified SNe are marked with brown ``x" markers, while the classified Type Ia and Type II SNe are marked with orange and pink stars, respectively. Dotted lines represent the population means. Top labels give the medians of each property for the BB galaxies, with error bars corresponding to the 16th and 84th percentiles.}
    \label{fig:property_corner}
\end{figure*}

\subsection{Comparing SED Parameters Derived from Spectroscopy}
\label{subsec:properties}

The SDSS BB sample is from DR13, which includes an abundance of spectral property measurements. Redshifts and spectral features are first determined by the SDSS spectroscopic pipeline \citep{stoughton_sloan_2002,albareti_13th_2017}. Subsequently, objects classified as `galaxy' are fitted with stellar population models to derive properties such as star formation rates, stellar masses, and metallicities \citep{chen_evolution_2012,maraston_stellar_2013,thomas_stellar_2013}.

In Figure~\ref{fig:property_corner}, we compare several intrinsic properties of the SN host galaxies to the full SDSS BB sample, which were assembled from the SDSS catalog by \citet{jiang_direct_2019}. We measure the difference between SN hosts and the broader BB population using several key physical parameters that are relevant to the production of ionizing photons in SFGs.  These include the H$\alpha$-derived SFR, the [O/H] metallicity, the SDSS-derived stellar mass from \citet{maraston_stellar_2013}, and the electron density from the flux ratio R $\equiv$ [SII]$\lambda$6716/[SII]$\lambda$6731 \citep{jiang_direct_2019}.

We find that non-Ia host galaxies tend to have higher SFR and stellar mass than the non-host BB population, with Kolmogorov–Smirnov (KS) statistics of 0.63 and 0.55, respectively, corresponding to p values of 0.03 and 0.08.  We also find offsets of $3.7\sigma$ (SFR) and $2.5\sigma$ (mass) significance in the difference between sample means (we note that we neglect modeling uncertainties in deriving these significances, assuming them to be subdominant). 

In contrast, the distributions of electron density and metallicity show no significant differences, with offsets at the $<1\sigma$ level. The elevated $H_{\alpha}$-derived SFRs of the SN hosts are consistent with expectations that CC\,SNe form from short-lived, massive stars and therefore occur preferentially in star-forming environments.
The shift toward higher stellar masses among SN hosts is likely due to the increase in size of the overall stellar population, meaning that more massive stars exist that can potentially explode as SNe, though higher stellar mass is also generally correlated with higher SFR in SFGs.

Though not included in our significance calculations, we also plot the SN\,Ia host galaxy in Figure~\ref{fig:property_corner}, which does not follow many of these trends. Due to the much longer delay times between star formation and explosion for many SNe\,Ia \citep{dilday_measurement_2008,maoz_observational_2014,keller_uncertainties_2022}, it is not surprising that weaker or different correlations between SNe\,Ia and galaxy properties may exist \citep[e.g.,][]{nugent_characterizing_2026}.

\subsection{UV Slopes}
\label{subsec:UV slopes}
The UV spectral slope \citep[usually measured within the window from $\sim1300--2500$ \AA;][]{calzetti_dust_1994}, $\beta$, is a tracer of both stellar age and dust content in SFGs \citep{reddy_mosdef_2015,chisholm_far-ultraviolet_2022}. Its strong correlation with the LyC escape fraction, $f_{\rm esc,LyC}$, is both theoretically predicted and demonstrated with data from studies such as the Low-Redshift Lyman Continuum Survey data \citep[LzLCS,][]{flury_low-redshift_II_2022, zackrisson_spectral_2013,saldana-lopez_low-redshift_2022,chisholm_far-ultraviolet_2022}.

We compute $\beta$ using Milky Way extinction-corrected GALEX {\it FUV} and {\it NUV} magnitudes for all BB and GP samples following Equation 1 of \citet[see also Equation 1 of \citealp{rogers_unbiased_2013}]{lin_discovery_2025}, which computes the slope from the photometric color:
\begin{equation}
\label{eqn:beta}
    \beta=-0.4\frac{m_1-m_2}{\log_{10}(\lambda_1/\lambda_2)}-2,
\end{equation}
where we use the corresponding filter pivot wavelengths $\lambda_{NUV}=2301$ \AA, and $\lambda_{FUV}=1535$ \AA. 

As a consistency check, we also compute $\beta$ using the SDSS $u$ and GALEX NUV colors, and find good agreement with the mean $\beta$ for the BB sample.\footnote{We use the subscript notations $\beta_{FUV-NUV}$ for GALEX color and $\beta_{\rm NUV-u}$ for the combination of GALEX {\it NUV} and SDSS $u$ band color.}  Between $\beta_{FUV-NUV}$ and $\beta_{\rm NUV-u}$, we measure a median absolute deviation that is consistent with the errors (i.e., a scatter of 0.275 after restricting the sample to $\beta_{\rm obs}$ errors on GALEX and SDSS to $<$0.2).  However, the SDSS $u$-band measurements are more uncertain as a whole, and for lower-redshift SNe ($z \lesssim 0.05$) they are potentially contaminated by [O\,II]$\lambda3727$ and $\lambda3729$ line flux (GALEX {\it FUV} may also be slightly contaminated by Ly$\alpha$ at $z > 0.11$ but Ly$\alpha$ would be at the very edge of the {\it FUV} bandpass where sensitivity is lower).  For this reason, we use the GALEX-computed $\beta_{FUV-NUV}$ slopes for this analysis.

\begin{figure}
    \centering
    \includegraphics[width=\linewidth]{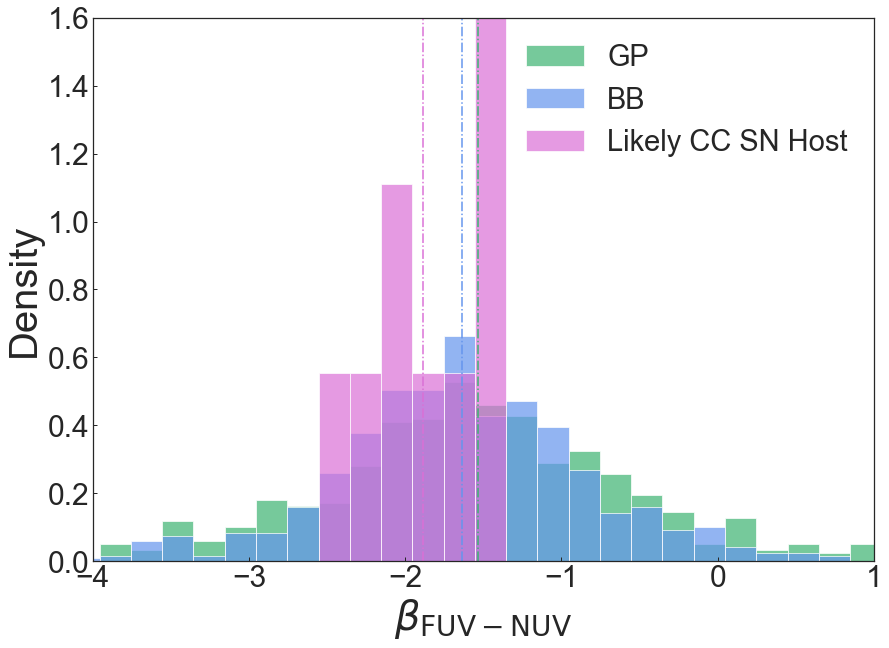}
    \caption{The UV slope, $\beta$, for GPs (green), BBs (blue), and BB SN hosts (pink).  The mean UV slope of SN hosts is lower than that of other BBs at 2.1$\sigma$ significance.  SN~2022emj (SN\,Ia) and SN~2020aho (blended photometry) are excluded from this figure.}
    \label{fig:beta_hist}
\end{figure}

We compare the distribution of $\beta$ between the BBs, GPs, and the SN hosts in Figure~\ref{fig:beta_hist}, and find that SN hosts show a lower mean UV slope, $\beta_{ FUV-NUV}=-1.88$ (median of $-1.89$), than the BB galaxies, which have an average slope of $-1.63$ (median of $-1.64$). The slopes differ at 2.1$\sigma$ significance, which we compute using the mean and standard error of the populations.
We exclude the SN\,Ia from these calculations, due to its long delay time, as well as SN\,2020aho due to substantial blending in GALEX filters with a nearby (and possibly interacting) companion galaxy.


\begin{figure}
    \centering
    \includegraphics[width=\linewidth]{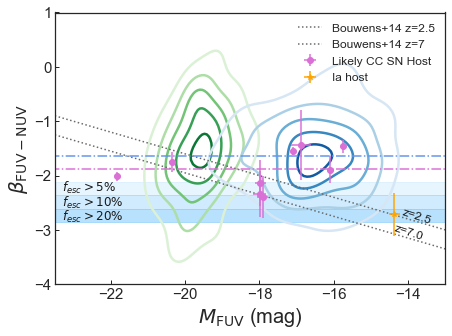}
    \caption{The $\beta$ parameter versus FUV magnitude ($M_{FUV}$) for GPs (green), BBs (blue), BB SN hosts (core-collapse, pink), and BB Type Ia host (orange star). SN~2020aho (blended photometry) is excluded from this figure. Estimated 5\%, 10\%, and 20\% LyC escape fractions are plotted using Equation 11 from \citet{chisholm_far-ultraviolet_2022}.  The best-fit $\beta$ versus $M_{FUV}$ relationships from \citet{bouwens_uv-continuum_2014} at $z=2.5$ and $z=7$ (grey dotted) are also shown, though we note that many BBs are significantly more UV-faint than the samples in this work.}
    \label{fig:beta_muv}
\end{figure}

In Figure~\ref{fig:beta_muv}, we compare $\beta$ to the GALEX {\it FUV} absolute magnitude, $M_{FUV}$. We overplot the median linear relationship between $\beta$ and $M_{FUV}$, fitted by \citet{bouwens_uv-continuum_2014} for $z=2.5$ and $z=7.0$. Although we do not observe any clear trends in the GP and BB populations, the SN host galaxies (pink) loosely follow the empirical relation of \citet{bouwens_uv-continuum_2014} and trend toward higher escape fractions for fainter host galaxies, though with significant scatter.

Because $\beta$ scales strongly with $f_{\rm esc,LyC}$, we can also use the analytic relation  proposed by \citet{chisholm_far-ultraviolet_2022} to convert $\beta$ to an approximate LyC escape fraction:
\begin{equation}
\label{eqn:f_esc}
    f_{\rm esc,LyC}=(1.3\pm0.6)\times10^{-4}\times10^{(-1.22\pm0.1)\beta}.
\end{equation}
We note that \citet{chisholm_far-ultraviolet_2022} fit for their UV continuum slope in the $1300-1800$\AA\ wavelength range, while the GALEX {\it NUV} filter extends redder, with an effective wavelength of $\sim$2300\AA. \citet{raiter_predicted_2010} show that measuring $\beta$ from 1800--2200\AA\ yields a higher $\beta$ measurement by $\sim$0.2--0.3~mag, and depends on the initial mass function and metallicity; however, in this work, the difference in $\beta$ compared to \citet{chisholm_far-ultraviolet_2022} is likely a more subtle $\sim$0.1--0.15~mag given that GALEX filters span a wide wavelength range from $\sim$1500--2300\AA.

Using this relation, we shade regions of Figure~\ref{fig:beta_muv} that correspond to escape fractions of $>5\%$, $>10\%$, and $20\%<f_{\rm esc,LyC}<40\%$. These show the number of GP/BB galaxies that are considered ``cosmologically relevant" ($>5\%$) as defined in \citet{chisholm_far-ultraviolet_2022}. We identify at least three out of ten CC\,SN hosts as likely having $f_{esc} > 5\%$. The host of the Type Ia SN (2022emj) may also have a LyC escape fraction of $\sim20\%$.

\begin{figure}
    \centering
    \includegraphics[width=\linewidth]{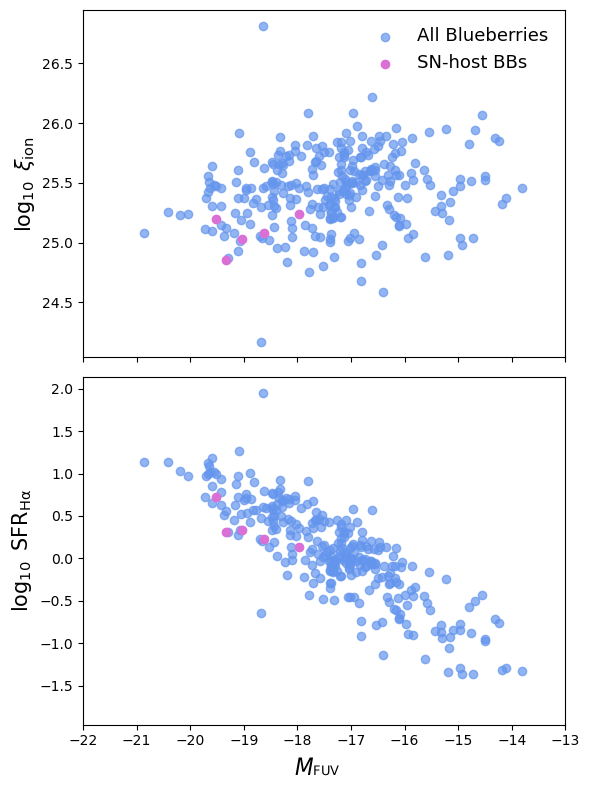}
    \caption{Comparisons between $\xi_{\rm ion}$, H$\alpha$-derived SFR, and {\it FUV} absolute magnitude for the SDSS BB sample (blue) and the SDSS SN-hosts (pink), excluding the host of the Type Ia SN~2022emj.
    The SN hosts have lower than average $\xi_{\rm ion}$ (top panel) due to their lower SFR when compared to BB galaxies with similar UV magnitudes (bottom panel).}
    \label{fig:xi_ion}
\end{figure}

Additionally, to assess whether the UV-slope distribution of the BB galaxies still differs from that of the SN hosts when controlling for the absolute UV magnitude of each galaxy --- given that SN hosts are brighter on average --- we perform a bootstrap test. We draw 10\,000 random subsamples of 10 non-host BB galaxies, each matched in $M_{\rm UV}$ to one of the SN-host galaxies. For each resample, we compute the median UV slope and compare it to that of the SN-host sample. Approximately $10\%$ of the resamples attain a median bluer than $-1.89$, the median UV slope of the SN-host galaxies. The average of the resampled medians is $-1.632$.

\subsection{Ionizing Photon Production}

The rate of ionizing photons leaked into the IGM for a given galaxy, $\Gamma$, is determined by both the fraction of ionizing photons that escape into the IGM ($f_{\rm esc}$) and the ionizing photon production efficiency $\xi_{\rm ion}$ \citep{emami_ionizing_2020}:
\begin{equation}
\label{eqn:ion_photon}
    \Gamma = \int L\Phi(L)\xi_{\rm ion}(L)f_{\rm esc}(L)dL.
\end{equation}
Here, $L$ is the galaxy luminosity and $\Phi(L)$ denotes the galaxy luminosity function.
In SFGs, the values of $\xi_{\rm ion}$ and $f_{\rm esc}$ can vary substantially depending on internal conditions such as ISM geometry and density structure \citep{fernandez_effect_2011,rinaldi_midis_2024}.

To quantify these properties in our samples, we compute $L_{H_{\alpha}}$ using the SDSS $H_{\alpha}$-derived SFR from \citep{kennicutt_past_1994} and use it to estimate $\xi_{\rm ion}$ following \citet{emami_ionizing_2020}:
\begin{equation}
\label{eqn:L_halpha}
    {\rm SFR}({\rm total})=\frac{L_{H_{\alpha}}}{1.26\times10^{41}~{\rm erg}\:s^{-1}}\;M_{\odot}\:{\rm yr}^{-1},
\end{equation}
\begin{equation}
\label{eqn:xi_ion}
    \xi_{\rm ion} = \frac{L_{H_{\alpha}}}{1.36\times10^{-12}L_{\rm UV}}~[s^{-1}/{\rm erg}\:s^{-1}\:Hz^{-1}].
\end{equation}

Figure \ref{fig:xi_ion} shows $\xi_{\rm ion}$ and the H$\alpha$-derived SFR as a function of {\it FUV} absolute magnitude for the subset of our CC\,SN hosts and BB galaxies with available SDSS H$\alpha$ measurements.  SN hosts tend to have brighter {\it FUV} magnitudes overall, which generally correspond to slightly lower $\xi_{ion}$; additionally, SN hosts appear to have lower H$\alpha$-derived SFR compared to other BBs of similar magnitude.  Compared to the full BB sample, we find that SN hosts exhibit lower $\xi_{\rm ion}$ values at $7.9\sigma$ significance.

We then apply the same bootstrap procedure used for $\beta$ in Section \ref{subsec:UV slopes} to our $\xi_{\rm ion}$ measurements to see whether SN hosts have lower $\xi_{\rm ion}$ when controlling for their FUV absolute magnitudes. Using $10\,000$ resamples of the BB population (each matched in size and UV magnitude to the SN-host sample), we find that only $2\%$ yield a median $\xi_{\rm ion}$ lower than that of the SN host galaxies.


To further understand the ionizing conditions of these populations, we also compare their  [O~III]/[O~II] ratio ([O~III]$\lambda\lambda4959,5007$/[O~II]$\lambda\lambda3726,3729$), again using SDSS measurements.  This is considered one of the optical diagnostics for the hard radiation content and ISM conditions, therefore providing additional information on the total production of ionizing photons \citep{jaskot_origin_2014,nakajima_ionization_2014,izotov_eight_2016,ma_binary_2016,ma_no_2020}. We find a SN hosts may have a lower O32 ratio at $1.8\sigma$ significance, consistent with our measurements of lower $\xi_{\rm ion}$.  

Overall, we note that the statistical significance of some of these differences is marginal given the small sample size; however, our finding that CC\,SN feedback predominantly occurs in UV-bright galaxies, and that these galaxies have lower $\xi_{ion}$ (by approximately $\sim$0.2--0.3 dex) is highly statistically significant.


\subsection{Star-formation Histories}
\label{subsec:SFH}

Finally, we use the {\tt Blast} tool from \citet{jones_blast_2024}\footnote{\url{https://blast.scimma.org/}.} to derive SED properties for the BB CC\,SN hosts, a randomly-selected subsample of BBs, and a randomly-selected subsample of field galaxies from SDSS (subsamples are necessary because computing SFHs for the full sample is computationally infeasible).  {\tt Blast} analyzes SN host properties by downloading archival images and automatically measuring matched-aperture photometry from archival SDSS \citep{york_sloan_2000}, the Dark Energy Spectroscopic Instrument Legacy imaging Surveys \citep{dey_overview_2019}, GALEX \citep{martin_galaxy_2005}, WISE \citep{wright_wide-field_2010}, Pan-STARRS \citep{chambers_pan-starrs1_2016,flewelling_pan-starrs1_2020,magnier_pan-starrs_2020}, and 2MASS images \citep{skrutskie_two_2006}.  It then uses the {\tt Prospector}-$\alpha$ model \citep{leja_how_2019,johnson_stellar_2021} to measure SED properties from this photometry, with a speed enhancement using simulation-based inference \citep{wang_sbi_2023}.  With the Prospector-$\alpha$ model, {\tt Blast} implements a non-parametric star-formation history across seven logarithmically spaced time bins.

\begin{figure}
    \centering
    \includegraphics[width=\linewidth]{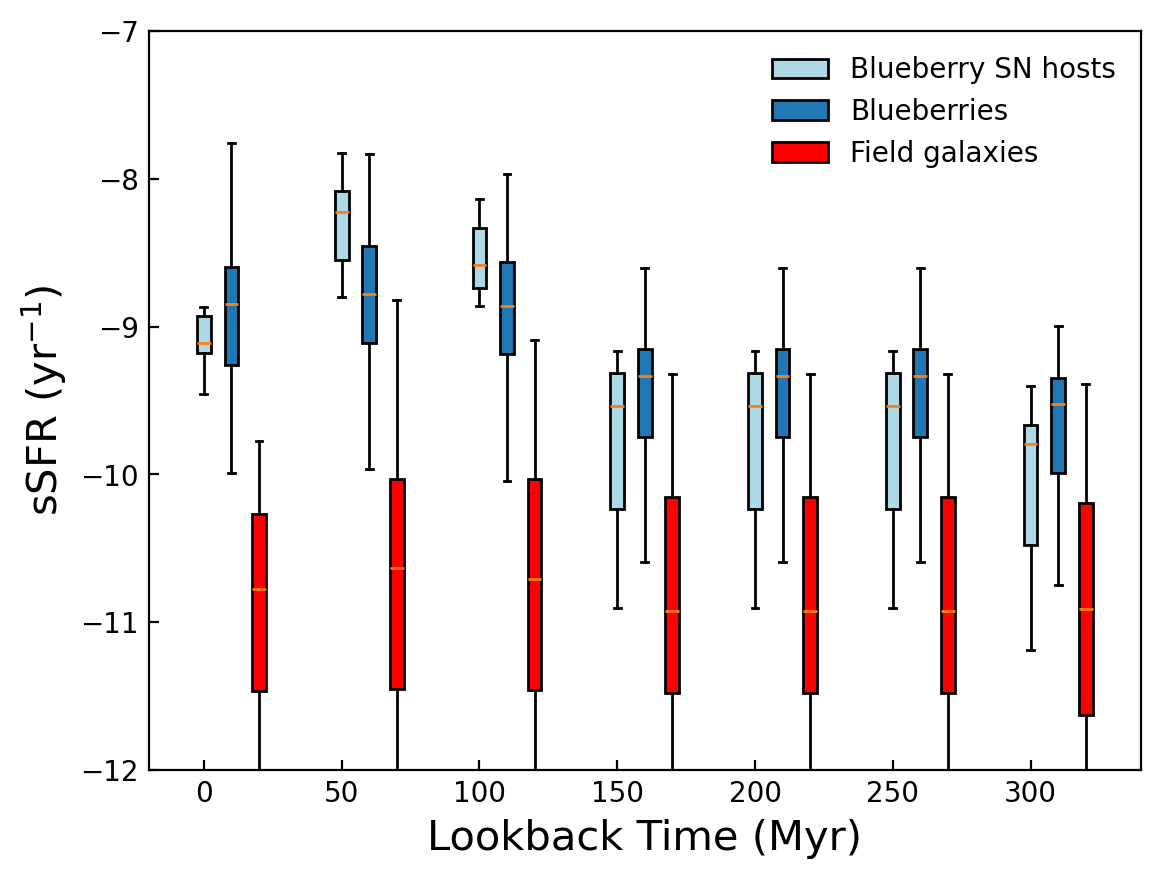}
    \caption{Box plot of the star formation history of SN hosts, BB galaxies, and SDSS field galaxies. The orange line represents the median SFR estimated by {\tt Blast}, while the box edges represent $25$ and $75$ percentiles of the population, and the whiskers for $\pm\:1.5\times$ the interquartile range, respectively. BB SN hosts show an elevated sSFR between $\sim30-100$ Myr, consistent with the CC SN delay time distribution \citep{zapartas_delay-time_2017,castrillo_delay_2021,saito_constraints_2022}. The SFH is computed from the prospector-$\alpha$ non-parametric model using bins of 0--30~Myr, 30--100~Myr, 100--300~Myr, and 300--1~Gyr.}
    \label{fig:sfh}
\end{figure}

Figure~\ref{fig:sfh} plots the SFH of the SN hosts, BB galaxies, and SDSS field galaxies. BB galaxies show an overall elevated SFR  compared to field galaxies. SN hosts specifically have a sSFR between lookback times of $\sim30$--$100$ Myr that is enhanced by 0.56~dex, and which is significant at the $99\%$ confidence level (using a Mann-Whitney Test).  This is consistent with expectations from the CC\,SN delay time distribution; the most massive SN progenitors (likely SN Ib or Ic) should explode in $\sim$10~Myr or less, while the least massive CC\,SN progenitors will be closer to 50~Myr between formation and explosion; this is consistent with the time frame of the observed starburst period. This spread in CC\,SN explosion timescales also allows time for changes in the ISM structure as a result of the injected mechanical energy, with $\sim2-3$~Myr typically needed for massive stars to clear out channels through the ISM via feedback winds and/or supernovae \citep{kimm_understanding_2019,naidu_synchrony_2022}.  However, the lower sSFR at the present time (as also seen in the lower panel of Figure \ref{fig:xi_ion}), is consistent with our findings of lower O32 ratios and $\xi_{\rm ion}$ in SN host galaxies.

\section{Discussion and Conclusions}
\label{sec:discussion}



In this study, we identified 11 SN events within 1242 $z < 0.12$ Blueberry (BB) galaxies. We find that SN host galaxies tend to exhibit elevated star formation rates, higher stellar masses, bursty star-formation histories, and have evidence for elevated UV escape fractions relative to non-host BBs. However, we also observe lower rates of ionizing photon production. Since GP and BB galaxies are widely regarded as analogs to early-universe Ly$\alpha$ emitters (LAEs), this gives us a unique tool for probing the role of SNe in facilitating ionizing photon escape during the EoR. 

\subsection{Implications for Reionization}

Our analysis of UV slopes, a proxy for $f_{\rm esc}$, finds modest evidence that BB galaxies that host SNe exhibit systematically steeper (more negative) UV slopes than the general BB population at 2.1$\sigma$ significance.
Bluer UV continua are often considered tracers of young stellar populations and/or reduced dust attenuation, both of which lead to higher ionizing photon escape fractions. In Figure~\ref{fig:beta_muv}, we further demonstrate that several SN host galaxies lie in regions that are considered ``cosmologically relevant" at $f_{\rm esc}^{\rm LyC}>5\%$, implying they would meaningfully contribute to the ionizing photon budget of cosmic reionization, based on the relation from \citet{chisholm_far-ultraviolet_2022}. Taken together, the steeper UV slopes and their alignment with LyC escape regions provide indirect but compelling evidence that SNe may contribute meaningfully to the escape of ionizing radiation, and potentially play a key role in EoR feedback processes.


Furthermore, these findings may be aligned with feedback-driven reionization models
\citep[e.g.,][]{jaskot_origin_2014,kimm_understanding_2019,flury_low-redshift_I_2022,flury_low-redshift_II_2022,naidu_synchrony_2022,flury_low-redshift_2025}, in which mechanical energy from SNe in stellar populations aged $\sim10$--$50$ Myr clears dense ISM and circumgalactic medium (CGM) regions. Our star-formation histories of SN hosts suggest a typical peak of SFR within the last $\sim$30--100~Myr, and ISM channels could form within 2-10~Myr of the first bursts of SNe \citep{naidu_synchrony_2022}. Starburst episodes in compact SFGs may last long enough for 
young stars to produce ionizing radiation that escapes via these SN-created channels \citep{izotov_green_2011,jaskot_origin_2014}.

However, the number of ionizing photons that ultimately escape from a galaxy also depends on its intrinsic ionizing photon production efficiency, $\xi_{\rm ion}$. 
We find that CC SN host galaxies exhibit lower $\xi_{\rm ion}$ compared to the non-host BB population with similar {\it FUV} absolute magnitudes, at 98\% confidence.
This implies that the intense burst of SFR that produced these SNe may begin to decrease on $\sim$50~Myr timescales, whether due to SN-driven feedback or other processes, and suggests that establishing the timescale of substantial ionizing photon production is 
also an important factor in understanding whether the contribution of SNe to reionization is truly significant.   
Studies on high-resolution radiation-hydrodynamic simulations have pointed out high variation in escape fractions in dwarf galaxies, again emphasizing the importance in understanding the timescales of LyC escape and ionizing photon production in these systems \citep{trebitsch_fluctuating_2017}. While the first SNe should clear channels on timescales as short as 10~Myr, our results align with previous delay time measurements and show that the average CC\,SN originates from older progenitors.




These results also offer potential insight into the relative roles of more massive systems within models of cosmic reionization, particularly in the context of democratic versus oligarchic scenarios. The SN-host BB galaxies in our sample have higher stellar masses than the full BB sample, with $\log(M_*/M_{\rm \odot}) \sim 8.5-9.5$ and a peak around $\sim 9.0$, consistent with the dominant contributors to the oligarchic reionization scenario \citep[log$(M_{*}/M_{\rm \odot})>8.5$,][]{naidu_rapid_2020}. 

In democratic reionization models, galaxies at the brighter and more massive end are often assumed to contribute minimally to the ionizing photon budget because their higher halo masses suppress sustained escape, with elevated $f_{\rm esc}$ occurring only during brief, SN-driven episodes \citep{finkelstein_conditions_2019}. However, our findings suggest that SN-host galaxies, despite being more massive than the commonly invoked low-mass reionizers, can nonetheless exhibit signatures consistent with elevated escape fractions (e.g., steeper UV slopes). This implies that SN-driven feedback may facilitate non-negligible ionizing photon escape even in deeper potential wells.

Although it is possible that SN feedback in more massive halos enables brief escape episodes, if CC SNe can be recurrent or clustered in highly SF galaxies, these episodes may also be repetitive and efficient \citep{mannucci_supernova_2005,anderson_multiplicity_2013}. Unlike in low-mass halos, where SN feedback may suppress star formation, in more massive systems, feedback may promote intermittent transparency without fully quenching star formation, allowing for sustained or repeated episodes of LyC leaking. This may be especially relevant for compact SFGs undergoing multiple, possibly overlapping starburst phases, potentially similar to GP and BB galaxies.  We also note that while SNe\,Ia are largely excluded from this analysis, they will also --- despite their low correlation with ongoing SFR --- inject additional, significant energy into the ISM of these galaxies, and should not be entirely neglected.


\subsection{Future Work}
This study is subject to several limitations. Most importantly, we are unable to directly observe the Ly$\alpha$ emission or LyC escape in these galaxies, as most only have optical spectra; we rely instead on indirect indicators such as UV slope and SFR estimates.  We also note that recent studies show that some low-mass/faint galaxies in the early Universe ($z\gtrsim6$) can have limited star-formation efficiencies, and that these early UV-faint galaxies can have vastly diverse SFHs \citep{endsley_burstiness_2025}. Therefore, whether local BBs are adequate analogs to the SFH of early galaxies is still a topic of debate.

Some of the observed trends, particularly the apparent suppression of SN rates in BB galaxies, may also be influenced by selection effects. For example, SNe occurring in very compact hosts can be systematically missed or misclassified; in particular, SNe in BB galaxies may be mistaken for AGN variability or otherwise blended with bright central emission. However, an intrinsic explanation cannot be ruled out: a reduced SN rate in compact galaxies could arise from differences in stellar populations, such as metallicity-driven mass function variations that suppress the formation of the most massive stars \citep{liang_luminosity_2024}. Distinguishing between these possibilities will require more complete and in-depth transient searches. Given the current limitations, we regard the present rate measurements as provisional and highlight the need for more systematic SN searches in compact, low-mass star-forming galaxies.


Future FUV data from instruments like the Hubble Space Telescope's (HST's) Cosmic Origins Spectrograph (COS) could more directly characterize Ly$\alpha$ emission and key ISM absorption diagnostics in these systems, including Si~II (a neutral gas tracer), C~II (cold gas), and C~IV (hot gas and outflows). Additionally, Ly$\alpha$ profile analysis (e.g., peak separations, asymmetry, and equivalent widths) could be used to directly characterize the covering fraction and column density of neutral hydrogen \citep{yang_green_2016,izotov_detection_2016}. Multiple past HST/COS programs have observed a subset of GP galaxies, even including a known SN host (2021bk, included in this study), with focus on their Ly$\alpha$ profiles and UV emission lines \citep{duval_lyman_2016,mcpherson_duvet_2023}.

Increased sample sizes and higher-redshift SN observations of GP galaxies will soon be possible, particularly from the Vera C. Rubin Observatory's Legacy Survey of Space and Time (LSST) and the {\it Roman Space Telescope}.
Simultaneously, high-redshift galaxy observations with {\it JWST} will increase the sample of high-$z$ Ly$\alpha$ measurements during the EoR, providing crucial insights into the SN-driven feedback mechanisms shaping the earliest galaxies \citep{jaskot_multivariate_2024,lin_quantifying_2024,ma_jwst_2024,dhandha_exploiting_2025,willott_search_2025}, and other drivers of cosmic reionization.

\section{ACKNOWLEDGEMENTS}
M.K.\ and D.O.J. acknowledge support from NSF grants AST-2407632, AST-2429450, and AST-2510993, NASA grant 80NSSC24M0023, and HST/JWST grants HST-GO-17128.028 and JWST-GO-05324.031, awarded by the Space Telescope Science Institute (STScI), which is operated by the Association of Universities for Research in Astronomy, Inc., for NASA, under contract NAS5-26555. N.E.D acknowledges support from NSF grants LEAPS-2532703 and AST-2510993.

Support for K.I. was provided by NASA through the NASA Hubble Fellowship grant HST-HF2-51508 awarded by the Space Telescope Science Institute, which is operated by the Association of Universities for Research in Astronomy, Inc., for NASA, under contract NAS5-26555.

B.C.L.\ is supported by the international Gemini Observatory, a program of NSF NOIRLab, which is managed by the Association of Universities for Research in Astronomy (AURA) under a cooperative agreement with the U.S. National Science Foundation, on behalf of the Gemini partnership of Argentina, Brazil, Canada, Chile, the Republic of Korea, and the United States of America.

W.B.H. acknowledges support from the National Science Foundation Graduate Research Fellowship Program under Grant No. 2236415.

The authors wish to recognize and acknowledge the cultural role and reverence that the summit of Maunakea has always had within the indigenous Hawaiian community.

%



\bibliography{references}



\end{document}